\documentclass[a4paper]{jpconf}
\usepackage{graphicx}
\usepackage{amsmath}
\usepackage{bm}
\usepackage{amssymb}
\usepackage[colorlinks=true,linkcolor=blue,citecolor=blue, urlcolor=blue]{hyperref}   % use for hypertext links, including those to external documents and URLs

\begin{document}
%\title{Fluctuating Hydrodynamics Confronts the Rapidity Dependence of Transverse Momentum Fluctuations}
\title{Rapidity Correlation Structures from Causal Hydrodynamics}

\author{Sean Gavin$^1$, George Moschelli$^2$, and Christopher Zin$^1$}

\address{$^1$Department of Physics and Astronomy, Wayne State University, 
Detroit, MI, 48202, USA\\
$^2$Lawrence Technological University, 21000 West Ten Mile Road, Southfield, MI  48075, USA}

\ead{
\href{mailto:sean.gavin@wayne.edu}{$^1$sean.gavin@wayne.edu}, 
\href{mailto:gmoschell@ltu.edu}{$^2$gmoschell@ltu.edu}, 
\href{mailto:aj3980@wayne.edu}{$^1$aj3980@wayne.edu}
}

%%%%%%%%%%%%%%%%%%%%%%%%%%%%%%%%%%%%%%%%%%%%%%%%%%%%%%%%%%%%%%%%%%%%%%%%%%%%%%%%%%%%%%%%%%%%%%%%%%%%%%%%%%%%%%%%%%%
%%%%%%%%%%%%%%%%%%%%%%% Abstract
%%%%%%%%%%%%%%%%%%%%%%%%%%%%%%%%%%%%%%%%%%%%%%%%%%%%%%%%%%%%%%%%%%%%%%%%%%%%%%%%%%%%%%%%%%%%%%%%%%%%%%%%%%%%%%%%%%%
\begin{abstract}
Viscous diffusion can broaden the rapidity dependence of two-particle transverse
momentum fluctuations. Surprisingly, measurements at RHIC by the STAR collaboration
demonstrate that this broadening is accompanied by the appearance of unanticipated structure
in the rapidity distribution of these fluctuations in the most central collisions. Although a first
order “classical” Navier-Stokes theory can roughly explain the rapidity broadening, it cannot
explain the additional structure. We propose that the rapidity structure can be explained
using the second order “causal” Israel-Stewart hydrodynamics with stochastic noise.

\end{abstract}

%%%%%%%%%%%%%%%%%%%%%%%%%%%%%%%%%%%%%%%%%%%%%%%%%%%%%%%%%%%%%%%%%%%%%%%%%%%%%%%%%%%%%%%%%%%%%%%%%%%%%%%%%%%%%%%%%%%
%%%%%%%%%%%%%%%%%%%%%%% SECTION: Introduction
%%%%%%%%%%%%%%%%%%%%%%%%%%%%%%%%%%%%%%%%%%%%%%%%%%%%%%%%%%%%%%%%%%%%%%%%%%%%%%%%%%%%%%%%%%%%%%%%%%%%%%%%%%%%%%%%%%%
\section{Introduction}\label{sec:intro}
Many experimental measurements demonstrate the profound effect that initial state fluctuations of nuclear collisions have on the subsequent dynamics of the system, but there is growing awareness that fluctuations and their dissipation also occur throughout the evolution. In earlier work we suggested that viscous diffusion broadens the rapidity dependence of two-particle transverse momentum correlations \cite{Gavin:2006xd}. That work stimulated an experimental analysis by STAR \cite{Agakishiev:2011fs} and the discovery that the rapidity structure of these correlations not only broadens but also develops a distinctively non-Gaussian structure not predicted by ref. \cite{Gavin:2006xd}. 

We employ second order hydrodynamics with stochastic noise to develop an evolution equation for two particle transverse momentum correlations \cite{Gavin:2016hmv}. In this work we will compare rapidity correlation structures evolving from first and second order diffusion. In sec. \ref{sec:theory} we briefly derive evolution equations for the transverse momentum correlation function as well as describe how this correlation function relates to experimental observables. In the interest of brevity, we leave the rigorous derivations to \cite{Gavin:2016hmv}. Instead we focus on the results in sec. \ref{sec:results}. We find that second order hydrodynamics is required to explain the measured rapidity correlation structure and consequently that these measurements can constrain a transport coefficient that is sensitive to the thermalization process. 

%%%%%%%%%%%%%%%%%%%%%%%%%%%%%%%%%%%%%%%%%%%%%%%%%%%%%%%%%%%%%%%%%%%%%%%%%%%%%%%%%%%%%%%%%%%%%%%%%%%%%%%%%%%%%%%%%%%
%%%%%%%%%%%%%%%%%%%%%%% SECTION: Fluctuations and Correlations
%%%%%%%%%%%%%%%%%%%%%%%%%%%%%%%%%%%%%%%%%%%%%%%%%%%%%%%%%%%%%%%%%%%%%%%%%%%%%%%%%%%%%%%%%%%%%%%%%%%%%%%%%%%%%%%%%%%
\section{Fluctuations and Correlations}\label{sec:theory}
Nuclear collisions produce a high energy density fluid that flows outward with an average transverse velocity $v_r$. Small deviations of the flow occur in each event. These deviations perturb the transverse momentum current of the fluid by an amount $g_t = T_{0r} - \langle T_{0r}\rangle$, where $T^{\mu\nu}$ is the stress energy tensor. Viscous friction arises as neighboring fluid elements flow past one another. This friction reduces velocity fluctuations, driving the velocity toward $v_r$ and the momentum deviations $g_t$ to zero. The final size of the fluctuations depends on the magnitude of the viscosity and the lifetime of the fluid.

We consider fluctuations of a fluid at rest with energy density $e$ and pressure $p$. Small fluctuations produce a small velocity $\mathbf{v}$ corresponding to a momentum current $\mathbf{M} \approx (e+p)\mathbf{v}$. To linear order in the fluctuations, we write the conservation form of the relativistic Navier-Stokes equation:
%
%%%%%%%%%%%%%%%%%%%%%%% EQUATION: Navier-Stokes
%
\begin{equation}\label{eq:NavStokes2}
 \frac{\partial}{\partial t} \mathbf{M}   + \bm{\nabla} p = \frac{\zeta + \tfrac{1}{3}\eta}{w}\bm{\nabla}(\nabla\cdot \mathbf{M}) + \frac{\eta}{w}\nabla^2\mathbf{M}
\end{equation}
where $\eta$ and $\zeta$ are the shear and bulk viscosity coefficients and $w=e+p$ is the enthalpy density. 
The momentum fluctuation density can be written in terms of curl free longitudinal modes $\mathbf{g}_l$ and divergence free shear modes $\mathbf{g}$ so we have $\mathbf{M} = \mathbf{g}_l+ \mathbf{g}$, where $\mathbf{\nabla} \times \mathbf{g}_l=0$ and $\mathbf{\nabla} \cdot \mathbf{g}=0$. By taking the curl of (\ref{eq:NavStokes2}) we find the shear modes satisfy 
%
%%%%%%%%%%%%%%%%%%%%%%% EQUATION: First order diffusion of shear modes
%
\begin{equation}\label{eq:DiffModes}
 \frac{\partial }{\partial t}\mathbf{g}   = \nu\nabla^2 \mathbf{g},
\end{equation}
where $\nu = \eta/w$ is the kinematic viscosity. Equation (\ref{eq:DiffModes}) shows that Fick's Law holds for the density of transverse momentum fluctuations from local equilibrium. 

We look for the diffusion of \textit{correlations} of the fluctuations $\mathbf{g}$ at different points in the fluid 
$r_g = \langle\mathbf{g}_1\mathbf{g}_2\rangle - \langle\mathbf{g}_1\rangle\langle\mathbf{g}_2\rangle $.
%
%%%%%%%%%%%%%%%%%%%%%%% EQUATION:  r_g correlation function
%
In equilibrium, thermodynamic noise will cause such correlations to occur so $r_g^{le}\neq 0$ and the interesting correlations are those that differ from the equilibrium value $\Delta r_g = r_g - r_g^{le}$. It turns out that for two particle momentum correlations (\ref{eq:DiffModes}) becomes $\left[\frac{\partial }{\partial t} - \nu\left(\nabla^2_1 + \nabla^2_2\right)\right] \Delta r_g = 0$. For a detailed derivation please see ref. \cite{Gavin:2016hmv}.

We assume the event-averaged flow velocity has the Bjorken form, $u^\mu=(t/\tau,0,0,z/\tau)$, where $\tau =(t^2 -z^2)^{1/2}$ is the proper time and $\eta = (1/2) \log ((t+z)/(t-z))$ is spatial rapidity.
In an expanding system, observe that the rapidity density of total momentum $G\equiv \int  g \tau dx_{\bot}$, where the integral is over the transverse area of the two colliding nuclei. If one identifies spatial
rapidity $\eta$ with the momentum-space rapidity of particles, then $G$ is observable and we look for correlations of the form
%
%%%%%%%%%%%%%%%%%%%%%%% EQUATION:  r_g correlation function
%
\begin{equation}\label{eq:RGdef}
r_G = \langle G(\mathbf{x}_1) G(\mathbf{x}_2)\rangle - \langle G(\mathbf{x}_1)\rangle\langle G(\mathbf{x}_2)\rangle
\end{equation}
Additionally we switch coordinates to relative $\eta_r = \eta_1 -\eta_2$ and average $\eta_a = (\eta_1 + \eta_2)/2$ rapidity and finally we have our first order result
%
%%%%%%%%%%%%%%%%%%%%%%% EQUATION: First order diffusion of r_g
%
\begin{equation}\label{eq:DiffModes2}
 \left[\frac{\partial }{\partial \tau} - 
 \frac{\nu}{\tau^2}\left( 2\frac{\partial^2}{\partial \eta_r^2} + \frac{1}{2}\frac{\partial^2}{\partial \eta_a^2}\right)
 \right] \Delta r_G = 0.
\end{equation}

Second order hydrodynamics is especially important for diffusive phenomena, where it renders the theory causal. In first order diffusion (\ref{eq:DiffModes}), a delta function perturbation instantaneously spreads into a Gaussian, with tails extending to infinity. New transport coefficients at second order include relaxation times for shear and bulk stresses, among other terms. Linearized forms of the second order equations are discussed in \cite{Romatschke:2009im,Young:2014pka}. To linear order the shear modes satisfy a Maxwell-Cattaneo equation
%
%%%%%%%%%%%%%%%%%%%%%%% EQUATION: Maxwell-Cattaneo 
%
\begin{equation}\label{eq:2DiffModes}
\left(\tau_\pi \frac{\partial^2}{\partial t^2} + \frac{\partial }{\partial t}\right)\mathbf{g}   = \nu\nabla^2 \mathbf{g}
\end{equation}
where the transport coefficient $\tau_\pi$ is a relaxation time for the shear modes. As with the first order case, we look to solve (\ref{eq:2DiffModes}) for the correlations (\ref{eq:RGdef}). We find 
%
%%%%%%%%%%%%%%%%%%%%%%% EQUATION: second order diffusion equation for Delta r_g   
%
\begin{equation}\label{eq:2VisDiff}
 \left[ \frac{\tau_\pi}{2}\frac{\partial^2}{\partial \tau^2}
  +\frac{\partial}{\partial \tau}
  -\frac{\nu}{\tau^2}\left( 2\frac{\partial^2}{\partial \eta_r^2} + \frac{1}{2}\frac{\partial^2}{\partial \eta_a^2} \right) \right]
  \Delta r_G = 0.
\end{equation}
For a detailed derivation please refer to ref. \cite{Gavin:2016hmv}. Observe that (\ref{eq:2VisDiff}) reduces to the first order result (\ref{eq:DiffModes2}) when $\tau_\pi=0$. Further observe that (\ref{eq:2VisDiff}) has features of both wave and diffusion equations. As we will discuss in sec. \ref{sec:results}, at early times, fluctuation signals propagate as waves while at later times diffusion fills in the space between the wave fronts. It is precisely this behavior that delays the growth of the rapidity correlation structure in comparison to the first order case. 

The correlation function $\Delta r_G$ is observable by measuring the covariance
%
%%%%%%%%%%%%%%%%%%%%%%% EQUATION:  C definition
%
\begin{equation}\label{C0def}
  {\cal C} =  \langle N\rangle^{-2}\langle \sum_{a\neq b} p_{t,a}p_{t,b}\rangle -\langle p_t\rangle^2 
  = \langle N\rangle^{-2}\int  \Delta r_G(\eta_r, \eta_a)d\eta_r d\eta_a,  
\end{equation}
where $a$ and $b$ label particles from each event, and the brackets represent the event average. The average transverse momentum is $\langle p_t\rangle \equiv \langle \sum_a p_{t,a}\rangle/\langle
N\rangle$. In the absence of correlations ${\cal C} = 0$, as is the case for local equilibrium in an infinite system.

The STAR collaboration at RHIC reports a differential version of the quantity $\cal C$ as a function of relative pseudorapidity $\eta_r$ and azimuthal angle $\phi_r$ of pairs: 
%
%%%%%%%%%%%%%%%%%%%%%%% EQUATION: C defined as pt covariance
%
\begin{equation}\label{exptC}
{\cal C}(\eta_r, \phi_r) = \left({\langle N \rangle_1 \langle N \rangle_2}\right)^{-1}
\left\langle \sum\limits_{a\neq b} p_{_{t,a}}p_{_{t,b}}\right\rangle_{1,2}
- \langle p_t\rangle_1 \langle p_t\rangle_2,
\end{equation}
where the numbers $\langle N \rangle_k$ and $\langle p_t \rangle_k$ refer to the particle number and transverse momentum in $(\eta_k, \phi_k)$ bins for particles $k= 1,2$ \cite{Agakishiev:2011fs}. We will discuss this observable in detail in the next section.

%%%%%%%%%%%%%%%%%%%%%%%%%%%%%%%%%%%%%%%%%%%%%%%%%%%%%%%%%%%%%%%%%%%%%%%%%%%%%%%%%%%%%%%%%%%%%%%%%%%%%%%%%%%%%%%%%%%
%%%%%%%%%%%%%%%%%%%%%%% SECTION: Results
%%%%%%%%%%%%%%%%%%%%%%%%%%%%%%%%%%%%%%%%%%%%%%%%%%%%%%%%%%%%%%%%%%%%%%%%%%%%%%%%%%%%%%%%%%%%%%%%%%%%%%%%%%%%%%%%%%%
\section{Results}\label{sec:results}
We will now explore the rapidity dependence of transverse momentum fluctuations. In an earlier work we show that the relative rapidity width, $\sigma$, of (\ref{exptC}) can be related to the shear viscosity of the system using first order diffusion \cite{Gavin:2006xd}. Here we discuss how the diffusion equation derived from second order hydrodynamics, (\ref{eq:2VisDiff}), modifies this result. In particular we demonstrate that the first order theory cannot simultaneously explain both the width and shape of (\ref{exptC}) in relative rapidity; where the first order theory fails, the second order theory succeeds. Interestingly, while the kinematic viscosity $\nu=\eta/Ts$ remains the dominant factor in determining the width, the relaxation time of the second order theory, $\tau_\pi$, is the primary influence on the shape.

\begin{figure}
\includegraphics[width=0.58\textwidth]{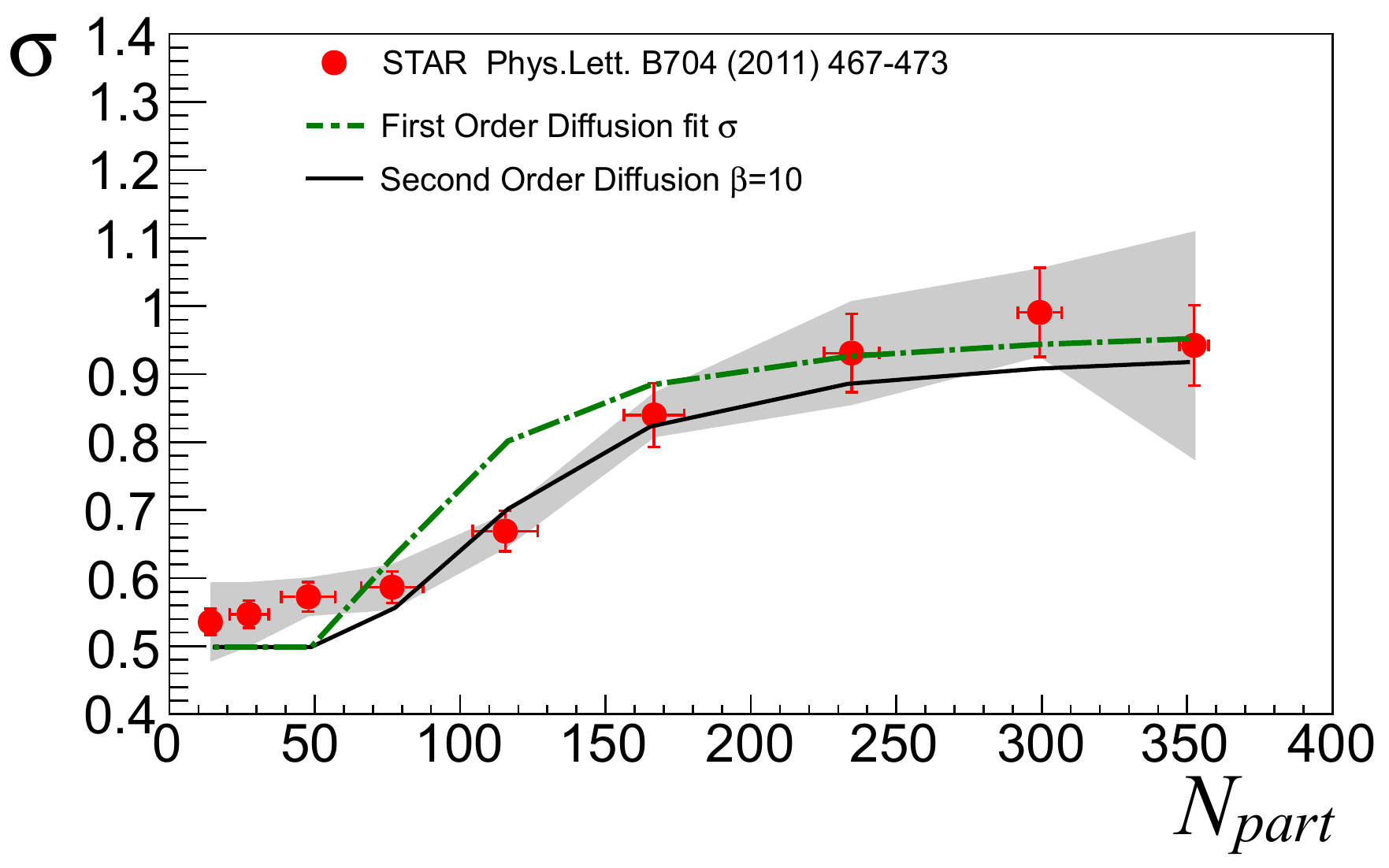}%
\hspace{0.02\textwidth}%
\begin{minipage}[b]{0.4\textwidth}\caption{\label{fig:sigNp}Rapidity width of ${\cal C}(\eta_r)$ as a function of the number of participants for second order momentum diffusion calculations (solid curve) compared to first order results (dot-dashed curve). Data (solid circles) from STAR include shaded area to denote the systematic uncertainty in the fit procedure \cite{Agakishiev:2011fs}. }
\end{minipage}
\end{figure}

\begin{figure}
\begin{center}
\includegraphics[width=0.75\textwidth]{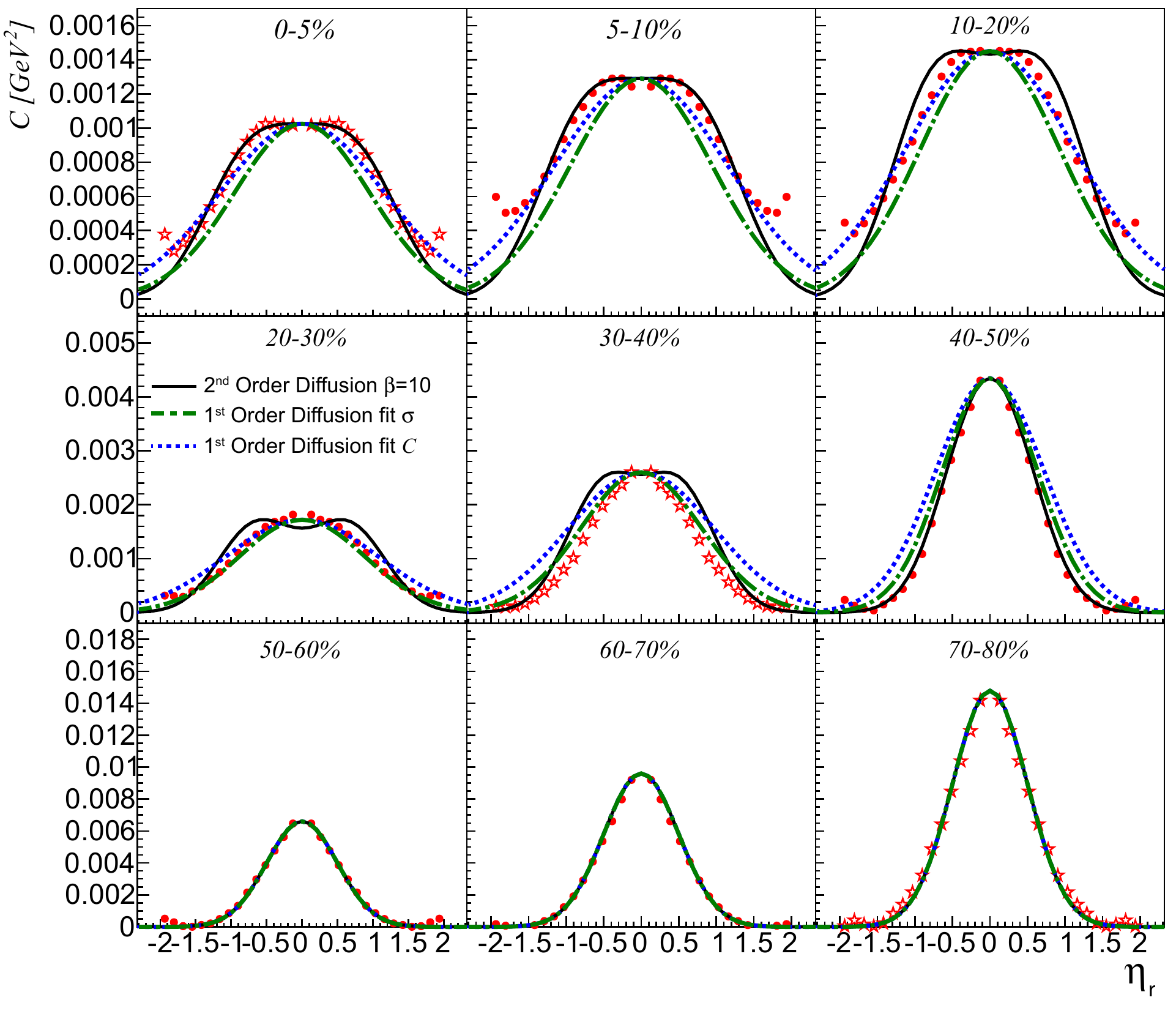}%
\end{center}
\caption{\label{fig:C12} Second order momentum diffusion calculations (solid curves) compared to the rapidity dependence of the measured covariance (\ref{exptC}). First order calculations are also compared for  best fit to $\sigma$ in fig.\ \ref{fig:sigNp} (dot-dashed curves) and the best fit to ${\cal C}(\eta_r)$ (dashed curves). Data (open stars) are from \cite{Agakishiev:2011fs} and (filled circles) from \cite{PrivComm}. Percentages of the cross section indicate centrality, with each panel corresponding to a width measurement in fig.\ \ref{fig:sigNp}. }
\end{figure}

In Au+Au collisions at the top RHIC energy, STAR has measured ${\cal C}$ in both its differential form (\ref{exptC}) and its relative rapidity width \cite{Agakishiev:2011fs}. 
The measured differential shape of (\ref{exptC}) resembles that of similar measurements of the ``ridge'' without $p_t$ weights. In Ref.\ \cite{Agakishiev:2011fs} authors interpret the near side, $|\phi_r| < 1$, of (\ref{exptC}) as a peak sitting on a flat pedestal and fit the $\phi_r$ integrated profile ${\cal C}(\eta_r)$ to extract the magnitude of the pedestal.
The measured width of ${\cal C}(\eta_r)$ is shown in fig.\ \ref{fig:sigNp}. Solid circles represent the RMS width in $\eta_r$ of (\ref{exptC}) integrated over an azimuthal range of $|\phi_r| < 1$ radians.
The gray error band in fig.\ \ref{fig:sigNp} represents the uncertainty in their fit procedure.
Figure \ref{fig:C12} shows the ${\cal C}(\eta_r)$ profile from which these widths are calculated. Open stars are from Ref.\ \cite{Agakishiev:2011fs} and solid circles are from Ref.\ \cite{PrivComm}. In all cases we have subtracted the pedestal.

%
%%%%%%%%%%%%%%%%%%%%%%% FIGURE: sigma vs Npart
%

The flat pedestal might be interpreted as the long-range correlations usually associated with the ridge while shorter range correlations would influence the shape of the peak sitting on the pedestal. Our calculations here specifically address the short range correlations. As discussed in Ref.\ \cite{Gavin:2006xd}, the growth of the width of this peak from peripheral to central collisions can be used to extract the viscosity. To compute the width in the second order theory, we follow ref.\ \cite{Aziz:2004qu} and multiply (\ref{eq:2VisDiff}) by $\eta_r^n$. Next, we integrate over $\eta_r$ and $\eta_a$ and use $\int\eta_r^n\partial^2\Delta r_G/\partial\eta_r^2 = n(n-1)\int\eta_r^{n-2}\Delta r_G$, which is nonzero only for $n \ge 2$.  We find 
%
%%%%%%%%%%%%%%%%%%%%%%% EQUATION: eta_r moments
%
\begin{equation}\label{eq:moments}
\left( \frac{\tau_\pi}{2}\frac{d^2}{d\tau^2} + \frac{d}{d\tau}\right)
A\langle\eta_r^n\rangle =
\frac{2\nu}{\tau^2}n(n-1)A\langle\eta_r^{n-2}\rangle, 
\end{equation}
where $\langle\eta_r^n\rangle = A^{-1}\int\eta_r^n\Delta r_G d\eta_r d\eta_a$ are the normalized moments of the rapidity correlation function. The amplitude $A$ and the mean $\langle \eta_r\rangle$ both satisfy (\ref{eq:moments}) with the  right side equal to zero. The vanishing first moment $\langle\eta_r\rangle=0$ makes physical sense if we assume a symmetric system. Therefore, the width can then be described as $\sigma^2 = \langle\eta_r^2\rangle - \langle\eta_r\rangle^2 = \langle\eta_r^2\rangle$ and following from (\ref{eq:moments}) we obtain the differential equation for the width,
%
%%%%%%%%%%%%%%%%%%%%%%% EQUATION: width differential equation
%
\begin{equation}\label{eq:moment2}
\left( \frac{\tau_\pi}{2}\frac{d^2}{d\tau^2} + \frac{d}{d\tau}\right)
\sigma^2 = \frac{4\nu}{\tau^2}.
\end{equation}

The solution of (\ref{eq:moment2}) yields the time dependence of the broadening of ${\cal C}(\eta_r)$. We can observe that we should expect that longer lived collisions have broader rapidity distribution by examining the first order case where $\tau_\pi=0$. The solution of (\ref{eq:moment2}) is then
%
%%%%%%%%%%%%%%%%%%%%%%% EQUATION: first order width
%
\begin{equation}\label{eq:DeltaV}
  \sigma^2 =\sigma_0^2+ \frac{4\nu}{\tau_0}\left(1-\frac{\tau_0}{\tau}\right),
\end{equation}
% \begin{minipage}{0.5\textwidth}
% %
%   \begin{equation}\label{eq:DeltaV}
%   \sigma^2 =\sigma_0^2+ \frac{4\nu}{\tau_0}\left(1-\frac{\tau_0}{\tau}\right),
%   \end{equation}
% %
% \end{minipage}
% %
% \begin{minipage}{0.5\textwidth}
% %
%   \begin{equation}\label{eq:DeltaVinf}
%   \sigma^2_\infty =\sigma_0^2+ \frac{4\nu}{\tau_0},
%   \end{equation}
% %
% \end{minipage}
% %
% ~\\
as found in Ref. \cite{Gavin:2006xd}. Here $\nu=\eta/Ts=const$ is the kinematic viscosity, $\tau_0$ is the formation time, and $\sigma_0$ is the width in peripheral collisions. STAR used this method to estimate the average shear viscosity to entropy density ratio to be $\eta/s=0.13\pm 0.03$ \cite{Agakishiev:2011fs}. 

To find the rapidity width for second order diffusion using (\ref{eq:moment2}) we must specify an initial condition for $d\sigma^2/d\tau\equiv \theta_0^2$ at $\tau = \tau_0$, the value of which is unknown.
We examine two cases, one where $\theta_0^2=0$ which assumes that no modifications to the correlation function (\ref{eq:RGdef}) occur prior to the formation time, and one that attempts to account for the effects of initial state expansion. In this proceedings we focus on the latter case and direct the reader to Ref. \cite{Gavin:2016hmv} for a detailed comparison of these two conditions.
For (\ref{eq:2VisDiff}), we take the initial correlation function to satisfy 
\begin{equation}\label{eq:CausalIC}
\frac{\partial \Delta r_G}{\partial \tau}\Big|_{\tau=\tau_0}
= \frac{\nu_0}{\tau_0^2}\left( 2\frac{\partial^2}{\partial \eta_r^2} + \frac{1}{2}\frac{\partial^2}{\partial \eta_a^2} \right)
  \Delta r_G,
\end{equation}
which corresponds to $\theta_0^2=4\nu/\tau_0^2$ in (\ref{eq:moment2}). Using (\ref{eq:CausalIC}) and solving (\ref{eq:moment2}) we find
%
%%%%%%%%%%%%%%%%%%%%%%% EQUATION: second order width
%
\begin{equation}\label{eq:SOwidth}
\sigma^2 = \sigma_0^2 +
\frac{\theta_0^2 \tau_\pi}{2} \left(1 - e^{-2(\tau-\tau_0)/\tau_\pi}\right)
+ \frac{8\nu}{\tau_\pi}\int\limits_{\tau_0}^\tau \! du \!
\int\limits_{\tau_0}^u \! \frac{ds}{s^2} e^{2(s-u)/\tau_\pi}.
\end{equation}
The solid black curve in fig.\ \ref{fig:sigNp} shows the value of (\ref{eq:SOwidth}) at the freeze out time parameterized as
%
%%%%%%%%%%%%%%%%%%%%%%% EQUATION: freeze out time
%
\begin{equation}\label{eq:tauF}
 \tau_F - \tau_0 = K(R(N_{part})-R_0)^2
\end{equation}
where $\tau_0$ is the formation time and $R_0$ is roughly the proton size. We compute $N_{part}$ and $R$ from a Glauber model and fix the constant $K$ so that the freeze out time in the most central collisions has a specified value $\tau_{Fc}$. 
Again we take $\nu = \eta/Ts=const$ for $\eta/s = 1/4\pi$, but now with $T=150$~MeV for all centralities. We must now specify the second-order relaxation time $\tau_\pi = \beta \nu$, for which we take $\beta = 10$.  The values $\tau_0 = 1.0$~fm and $\tau_{Fc} = 10$~fm then give superb agreement with data. 

The dash-dot curve shows our best fit to this data using the first order result (\ref{eq:DeltaV}) evaluated at $\tau_F$, eq.\  (\ref{eq:tauF}).  Again $\eta/s = 1/4\pi$ but we take the freeze out temperature to be $T = 140$~MeV for all centralities. Values of the space time parameters $\tau_0 = 0.65$~fm and $\tau_{Fc} = 12$~fm then specify the lifetime (\ref{eq:tauF}).   
One might argue the rough agreement in fig.\ \ref{fig:sigNp} for the first order result is compelling were it not for the fact that it consistently overestimates the data in the region where the data grows the fastest.
It is a feature of first order diffusion that it occurs a-causally; correlations diffuse too rapidly. Hence, we see that collision systems that freeze out at earlier times disagree with the first order case while more central, longer lived collision systems match with the asymptotic limit of (\ref{eq:DeltaV}). For this reason the estimate of the shear viscosity in Ref. \cite{Agakishiev:2011fs} is still on strong footing since authors used only the difference in central and most peripheral widths.

%
%%%%%%%%%%%%%%%%%%%%%%% FIGURE: C vs tau first and second order
%
\begin{figure}
\begin{minipage}{0.485\textwidth}
\includegraphics[width=\textwidth]{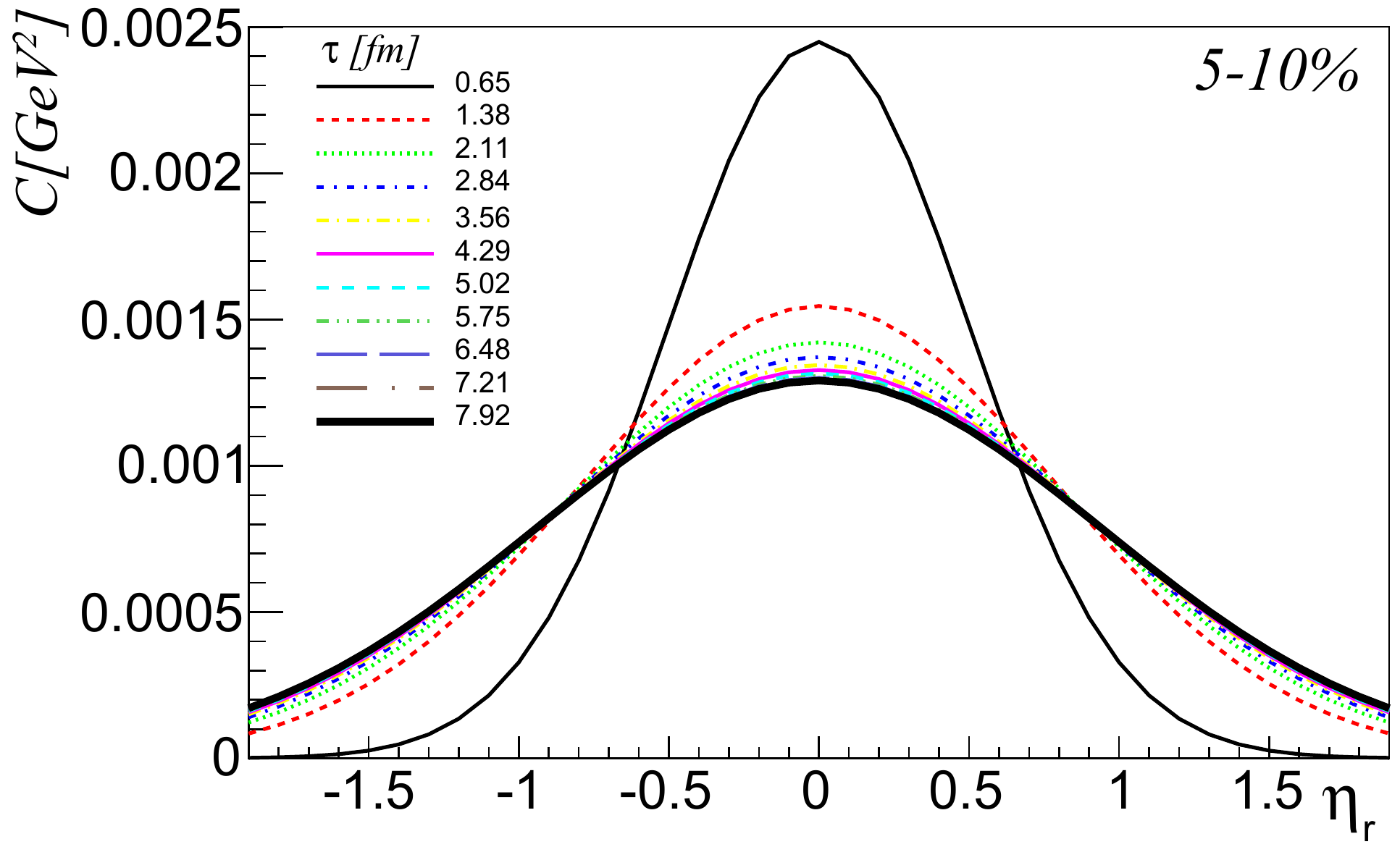}
\caption{\label{fig:Ctau1st}Time evolution of the rapidity structure of transverse momentum correlations from first order diffusion (\ref{eq:DiffModes2}) for $5-10\%$ central collisions.}
\end{minipage}\hspace{0.03\textwidth}%
\begin{minipage}{0.485\textwidth}
\includegraphics[width=\textwidth]{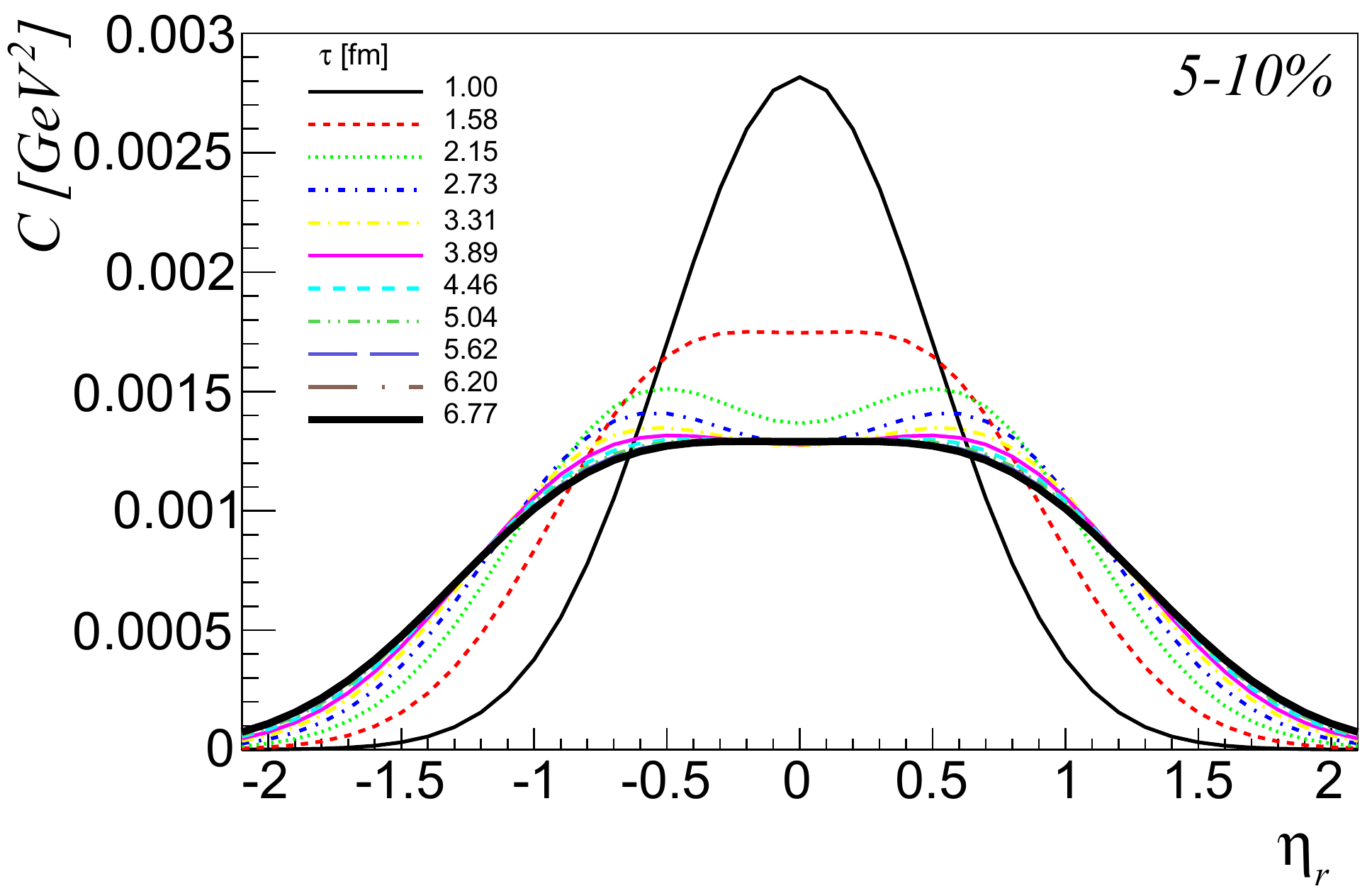}
\caption{\label{fig:Ctau2nd}Time evolution of the rapidity structure of transverse momentum correlations from second order diffusion (\ref{eq:2VisDiff}) for $5-10\%$ central collisions.}
\end{minipage} 
\end{figure}

The asymptotic value of the first order solution (\ref{eq:DeltaV}) is
%
%%%%%%%%%%%%%%%%%%%%%%% EQUATION: first order width asymptotic limit
%
\begin{equation}\label{eq:DeltaVinf}
\sigma^2_\infty =\sigma_0^2+ \frac{4\nu}{\tau_0}.
\end{equation}
This saturation of the rapidity width to the value (\ref{eq:DeltaVinf}) is a straightforward consequence of Bjorken flow. In a stationary liquid, a spike in momentum diffuses over a range $\sim (2\nu t)^{1/2}$ that grows with time $t$.  Bjorken expansion of the underlying fluid stretches the longitudinal scale $\propto t$, rapidly overtaking diffusion and ``freezing in'' the initial inhomogeneity. Correspondingly, one can observe this same behavior in (\ref{eq:2VisDiff}). At long times the rightmost term vanishes, meaning the rapidity of particles ceases to change due to shear forces.
Consequently, the rapidity width from second order diffusion, in the case of $\theta_0^2=0$, will reach the same asymptotic value. Importantly, the difference between the first and second order cases is that in the second order case the growth of the rapidity width, or equivalently the evolution of the correlation function (\ref{eq:RGdef}), is delayed according to the relaxation time $\tau_\pi$.

The freeze out time plays an important role in selecting the state of the correlation function measured by the experiment. For constant $\nu$ and $\tau_\pi$ all centralities follow the same trajectory in the solution of (\ref{eq:2VisDiff}) and consequently (\ref{eq:moment2}). If $\nu$ and $\tau_\pi$ change throughout the evolution, each collision would have a unique solution to (\ref{eq:2VisDiff}), but we leave that to future work. Figures \ref{fig:Ctau1st} and \ref{fig:Ctau2nd} show the time evolution of the correlation function for $5-10\%$ central collisions for the first and second order cases respectively. 
In both cases we assume the initial transverse momentum correlation function to be $\Delta r_G(\eta_r,\eta_a,\tau_0) = A e^{-\eta_r^2/2\sigma_0^2}e^{-\eta_a^2/2\Sigma_0^2}$. This distribution is motivated by the rapidity dependence of measured correlation functions for multiplicity and net charge in pp collisions. We set the initial width in relative rapidity, $\sigma_0$ to fit the most peripheral distribution in fig.\ \ref{fig:C12}. Furthermore, we assume there is insufficient time for significant evolution in the three most peripheral cases in fig.\ \ref{fig:C12}. The data supports this claim and gives a consistent value of $\sigma_0 = 0.50$. The average pseudo-rapidity width $\Sigma_0\approx 5-6$ units is assumed to be a ``large'' value relative to the size of experimental acceptance. We will take $A$ to fit the peak value of the measured $\cal{C}$. This parameter has little impact on our current study, since we are only concerned with the shape of the function.  We use (\ref{eq:CausalIC}) for the initial value of the first derivative.

%
%%%%%%%%%%%%%%%%%%%%%%% FIGURE: C vs Delta Eta
%

The striking difference between figures \ref{fig:Ctau1st} and \ref{fig:Ctau2nd}  is our primary result. In the first order case, diffusion is the only process taking place; starting from an initial Gaussian shape the rapidity distribution of correlations will always remain Gaussian. In the second order case, fluctuation signals propagate in opposite directions with a wave speed of $v=\sqrt{\nu/\tau_\pi}$ in Cartesian coordinates. In rapidity coordinates, as longitudinal expansion overtakes the signals, fluctuations are frozen into the final distribution. Depending on what time this takes place a flattened or even bimodal peak structure is visible. Indeed STAR's discovery of this type of rapidity structure in the measurement of (\ref{exptC}), shown in fig.\ \ref{fig:C12}, is the motivation for this work.

In fig.\ \ref{fig:C12} we show comparisons of first and second order cases to measured data (with the pedestal subtracted). The most noticeable feature is that in central collisions there is a flattening of the peak and even hints of a double bump structure. First order diffusion will never be able to achieve this shape regardless of choices of parameters. We illustrate this with two choices of parameter sets. For the dot-dashed curves we use the same parameters chosen to fit the width in fig. \ref{fig:sigNp}. Agreement is poor. Since the first order shape is always Gaussian, much of the value of the width will depend on how the tails are truncated to match the acceptance of the data. If we ignore the acceptance constraints and just try to match ${\cal C}(\eta_r)$ in fig. \ref{fig:C12} as best as possible we find the dashed curves for parameter values $\eta/s = 1/4\pi$, $T = 110$~MeV, $\tau_0 = 0.50$~fm, and $\tau_{Fc} = 10$~fm. Agreement with the measured shape is still quite poor. The solid black curves use the second order case with the same parameters as used in fig. \ref{fig:sigNp}. Agreement with data is much better than either of the first order cases and we see signs of the flattening of the peak in central collisions.

\begin{figure}
\includegraphics[width=0.58\textwidth]{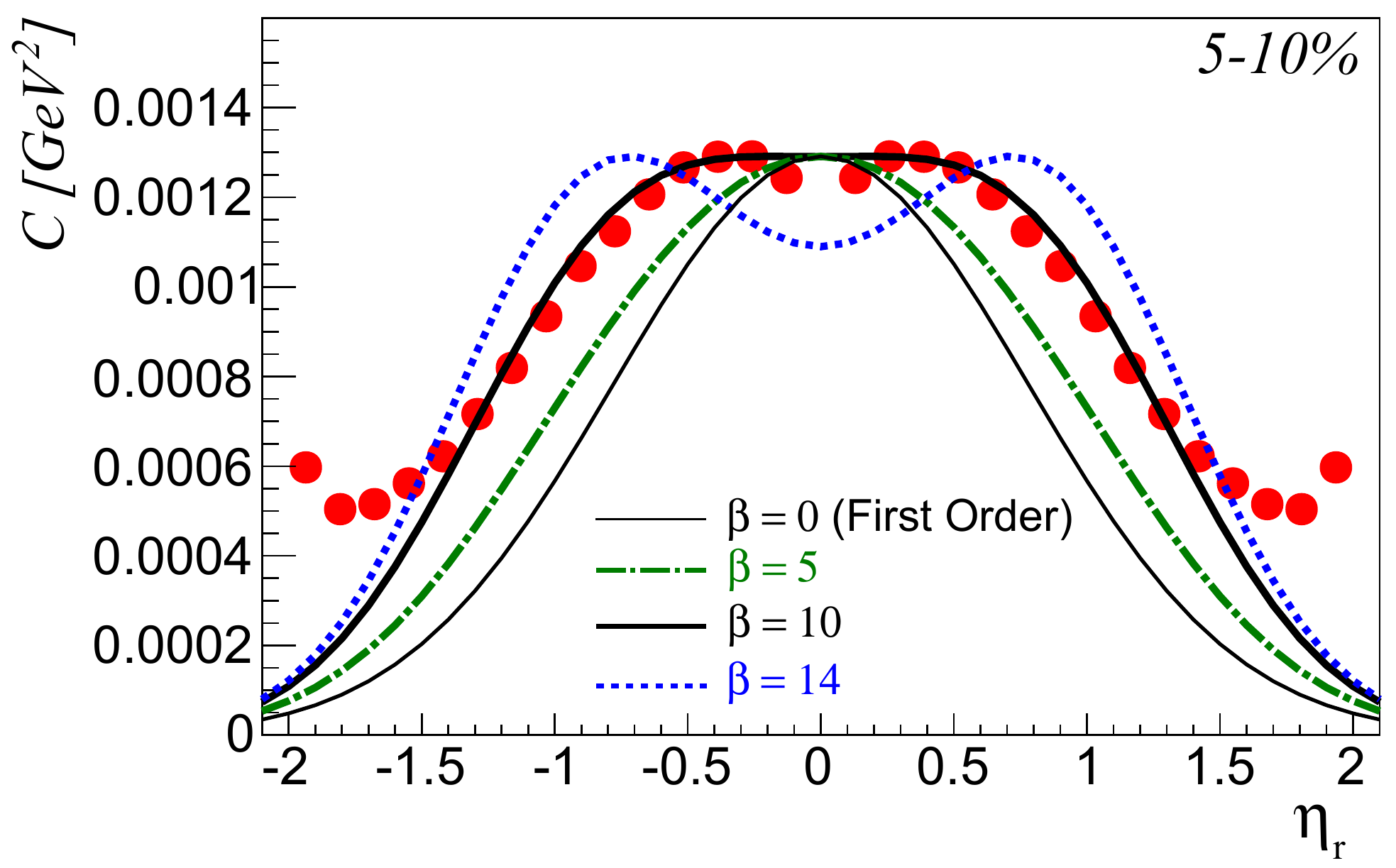}%
\hspace{0.02\textwidth}%
\begin{minipage}[b]{0.4\textwidth}\caption{\label{fig:ChangeBeta} Rapidity structure of transverse momentum correlations at freeze out from second order diffusion (\ref{eq:2VisDiff}). Different values of $\tau_\pi=\beta\nu$ are chosen by changing the parameter $\beta$. Results are selected for $5-10\%$ central collisions and compared to data \cite{Agakishiev:2011fs,PrivComm}. }
\end{minipage}
\end{figure}

The flattening of the peak is definitively a second order effect that is modulated by the relaxation time $\tau_\pi=\beta\nu$. This suggests that the experiment has access to this relaxation time. In fig. \ref{fig:ChangeBeta} we examine the effect of $\tau_{\pi}$ on our $5-10\%$ centrality result. In all cases we use the same parameter choices as used in in figs. \ref{fig:sigNp}, \ref{fig:Ctau2nd}, and \ref{fig:C12} with the exception of $\beta$. Previously for all second order results we chose $\beta=10$ since it yielded the most consistently strong agreement with all centralities. This solution is represented as the thick black line in fig. \ref{fig:ChangeBeta}. Incidentally, kinetic theory of massless Boltzmann particles predicts a value of $\beta=5$. Larger values of $\beta$ correspond to slower propagation of fluctuation signals in the medium, a fact that could be used to characterize the nature of the medium. The slower signal propagation associated with increasing $\beta$ means that the system takes longer to reach the asymptotic diffusion limit and longitudinal expansion ``freezes'' the correlation structure earlier in its development. As we can see in fig. \ref{fig:Ctau2nd}, at earlier times the wave nature of the structure still dominates the correlation shape. So, for larger $\beta$ values we should see an emergence of a more well defined double bump structure, as we see in the dashed line in fig. \ref{fig:ChangeBeta} corresponding to $\beta=14$. 

We remark that the resolution of the double peak structure also depends on the narrowness of the initial width $\sigma_0$. If the initial correlation structure is wide then we may never see a well defined double peak structure, but that structure will still be decidedly non-Gaussian. Conversely if the initial structure is very narrow we could still see a double peak structure with smaller $\beta$ values.

%
%%%%%%%%%%%%%%%%%%%%%%% FIGURE: C different betas on 0-5%
%
%
%%%%%%%%%%%%%%%%%%%%%%%%%%%%%%%%%%%%%%%%%%%%%%%%%%%%%%%%%%%%%%%%%%%%%%%%%%%%%%%%%%%%%%%%%%%%%%%%%%%%%%%%%%%%%%%%%%%
%%%%%%%%%%%%%%%%%%%%%%% SECTION: Conclusion
%%%%%%%%%%%%%%%%%%%%%%%%%%%%%%%%%%%%%%%%%%%%%%%%%%%%%%%%%%%%%%%%%%%%%%%%%%%%%%%%%%%%%%%%%%%%%%%%%%%%%%%%%%%%%%%%%%%
\section{Conclusions}\label{sec:conclusion}
In this paper we study the rapidity correlation structure of transverse momentum correlations in nuclear collisions (\ref{C0def}). In earlier work we suggested how these correlations could be used to study viscosity \cite{Gavin:2006xd}.  In an effort to obtain an estimate of the shear viscosity to entropy density ratio, STAR measured a differential version of these correlations (\ref{exptC}) and, in doing so, discovered that the rapidity shape has a distinctive non-Gaussian flattening in the more central centrality classes. In ref. \cite{Gavin:2016hmv} we use second order hydrodynamics with stochastic noise to develop a second order diffusion equation (\ref{eq:2VisDiff}) for the evolution of transverse momentum fluctuations that lead to the correlations (\ref{C0def}). 

Comparisons of first and second order diffusion results to data in figs. \ref{fig:sigNp} and \ref{fig:C12} suggest that first order diffusion cannot be used to simultaneously explain both the broadening of and the emergence of the non-Gaussian shape of the correlation structure from peripheral to central collisions. We attribute this failure to the a-causal nature of first order diffusion where fluctuation signals can propagate at infinite speeds throughout the medium. In second order diffusion fluctuation signals propagate as wave fronts with speed $\sqrt{\nu/\tau_\pi}$ where $\nu=\eta/Ts$ is the kinematic viscosity and $\tau_\pi=\beta\nu$ is the relaxation time related to the transport of the fluctuation signal through the medium. We solve the second order diffusion equation (\ref{eq:2VisDiff}) for constant $\nu$ and $\tau_\pi$ to find excellent agreement with data using a value of $\beta=10$. We further find that the kinematic viscosity is the primary factor in determining the broadening of the correlation structure and $\tau_\pi$ has significant influence over the flattening or even bimodal nature of the correlation peak.

The emergence of a flattened peak in the rapidity correlation structure and the natural way in which second order hydrodynamics accommodates this behavior suggests that experiments have the ability to measure the relaxation time $\tau_\pi$. Since $\tau_\pi$ is the characteristic time for the dissipation of fluctuations in the medium, studying this quantity can yield information on the thermalization process.

%%%%%%%%%%%%%%%%%%%%%%%%%%%%%%%%%%%%%%%%%%%%%%%%%%%%%%%%%%%%%%%%%%%%%%%%%%%%%%%%%%%%%%%%%%%%%%%%%%%%%%%%%%%%%%%%%%%
%%%%%%%%%%%%%%%%%%%%%%% SECTION: Acknowledgements
%%%%%%%%%%%%%%%%%%%%%%%%%%%%%%%%%%%%%%%%%%%%%%%%%%%%%%%%%%%%%%%%%%%%%%%%%%%%%%%%%%%%%%%%%%%%%%%%%%%%%%%%%%%%%%%%%%%
\ack%{Acknowledgements}
We thank Rajendra Pokharel for collaboration in the early stages of this work. Special thanks to Monika Sharma and Claude Pruneau for discussing the STAR data. We thank Victoria Drolshagen, Mauricio Martinez, Jaki Noronha-Hostler, Jorge Noronha, Scott Pratt, and Clint Young for discussions.  This work was supported in part by the U.S. NSF grant PHY-1207687.

\section*{References}

\bibliographystyle{utphys}
\bibliography{ptDiffusion_References}

\end{document}